\documentclass[final,5p,times,twocolumn]{elsarticle}

\usepackage{amssymb}

\usepackage{lineno}

\usepackage[multidot]{grffile} 
\usepackage[separate-uncertainty=true, multi-part-units=single]{siunitx}
\usepackage{hyperref} 

\usepackage{xcolor}   

\journal{Journal of Crystal Growth}

\begin{document}

\begin{frontmatter}



\title{The initial growth of sidebranches in ammonium chloride dendrites}


\author{A. J. Dougherty}

\address{Department of Physics, Lafayette College, Easton, PA 18042-1789, USA}

\begin{abstract}
We report measurements for the initial stages of
sidebranching during the dendritic growth of ammonium chloride
from supersaturated aqueous solution.  The earliest sidebranches
are approximately periodic; they are first evident about $36 \rho$
behind the tip, where $\rho$ is the tip radius, and have an average
initial spacing of about $5 \rho$, though both values show considerable
variation.  The initial sidebranch amplitude grows
approximately exponentially, but quickly saturates as sidebranches compete
and coarsening sets in.  This initial sidebranch growth is reasonably
consistent with what would be expected for noise-driven sidebranches,
though there are some quantitative differences.
\end{abstract}


\end{frontmatter}



\section{Introduction}

Dendritic crystal growth is frequently observed during the crystallization
of non-faceted materials from pure melt or supersaturated solution.
Common examples include structures formed during the solidification of
many metals and metal alloys \cite{Langer1980}.
For reviews, see Glicksman \cite{Glicksman2015-review}, Jaafar \textit{et
al.}\ \cite{Jaafar2017}, Asta \textit{et al.}\ \cite{Asta2009}, and
Boettinger \textit{et al.}\ \cite{Trivedi2000}.

Dendrites are also more conveniently observed in the crystallization
of transparent model compounds; see Akamatsu and Nguyen-Thi
\cite{Akamatsu2016} and Huang and Wang \cite{Huang2012}
for overviews.  Examples include succinonitrile and pivalic
acid \cite{Glicksman2015-review}, rare gases, such as helium
\cite{Rolley1994} and xenon \cite{Bilgram1993}, and some salts, such as
ammonium bromide \cite{Dougherty1987, Couder2005} and ammonium
chloride \cite{Honjo1985, Tanaka1992, Sawada1995, Dougherty2005, Martyushev2016}.

Dendritic crystals are characterized by an initially smooth tip that grows
at approximately constant speed.  Due to the anisotropy of the underlying
crystal structure, the tip is not axisymmetric, but instead develops
fins along the preferred growth directions.  Those fins are unstable, and
sidebranches begin to develop on top of them a short distance behind the
tip.  Much of the beauty of complex crystal structures results from the
intricate development and subsequent competition of those sidebranches.

Considerable theoretical and experimental effort has focused on
understanding the origin of the sidebranches,
measuring the properties of the sidebranch structure,
and looking for scaling laws that
might govern their ultimate development.  Previous extensive studies
of sidebranch structure have been reported for succinonitrile
\cite{Glicksman1981a, Corrigan1999a, Beckermann1998},
pivalic acid \cite{Giummarra2005a, Dougherty1994},
ammonium bromide \cite{Dougherty1987, Couder2005},
xenon \cite{Bisang1996, Wittwer2006},
ammonium chloride \cite{Dougherty1994, Dougherty1992, Kishinawa2008}, and
succinonitrile and succinonitrile-acetone alloys
\cite{Beckermann2012}.

In this paper, we focus on the early sidebranches in the dendritic
crystal growth of ammonium chloride at low supersaturation, and compare
the results to those predicted for noise-driven sidebranches.
Specifically, we consider three quantities:  the distance to the first
sidebranch, $z_\mathrm{sbr}$, the initial sidebranch spacing, $\lambda$,
and the amplitude of the initial sidebranch envelope, $A(z)$.  For
noise-driven sidebranches, predictions for these quantities were developed
by Langer for axisymmetric dendrites \cite{Langer1987}, and extended to
non-axisymmetric dendrites by Brener and Temkin \cite{Brener1995}.
Details are in the theory section below.

Some of these predictions have been tested for the dendritic growth of
pure materials.  For succinonitrile \cite{Glicksman1981a, Corrigan1999a},
the measured values of $z_\mathrm{sbr}$ and $\lambda$ are in accord
with the predictions of Ref.~\cite{Brener1995}.  Similarly, for xenon,
both the measured value for $z_\mathrm{sbr}$ \cite{Bisang1996} and the
sidebranch envelope $A(z)$ \cite{Wittwer2006} were found to be consistent
with the theory.  For pivalic acid \cite{Giummarra2005a}, on the other
hand, the measured value of $\lambda$ is larger than the predicted value.

This paper adds to the range of materials tested by examining the
dendritic growth of ammonium chloride from aqueous solution.  We find
that the measured value for $z_\mathrm{sbr}$ is consistent with the
theory in Ref.~\cite{Brener1995}, but the measured spacing $\lambda$
is smaller than expected.
The measured sidebranch amplitude $A(z)$ follows a modified version
of the predicted shape, but only for a narrow range of distances $z$.
The underlying causes for these differences remain unclear.

\section{Theory}

The basic background theory for steady state diffusion-limited dendritic crystal
growth is presented in \cite{Langer1980}.  Solutions incorporating anisotropy in
the surface energy, known as microscopic solvability, are presented in
\cite{Kessler1988}.  Briefly, for slow growth when kinetic effects may be
ignored, the crystal is characterized by
a smooth tip of radius $\rho$ growing at constant
speed $v$.  The tip radius and velocity are related to the
dimensionless ``stability constant'' $\sigma ^*$ by
\begin{equation}
\sigma^* = \frac{2 d_0 D}{v \rho^2} ,
\label{eqn:sigmastar}
\end{equation}
where $D$ is the relevant diffusion constant, and $d_0$ is the
capillary length, which is related to the solid-liquid interface energy.
The predicted value for $\sigma^*$ depends on the anisotropy in
the surface energy.

Direct testing of Eq.~\ref{eqn:sigmastar} has proven difficult, in
part due to challenges in measuring all of the relevant materials
properties to sufficiently high precision, and in part due to the
complications introduced by convection in most terrestrial experiments.
Moreover, in both microgravity \cite{Glicksman1995a, Lacombe2007a}
and terrestrial experiments \cite{Beckermann2012, Martyushev2016},
as well as in phase-field numerical simulations \cite{Mullis2011,
Bollada2015}, the value for $\sigma^*$ is not constant, but decreases
with increasing undercooling or supersaturation over reasonable ranges
of experimental interest.  Nevertheless, it still provides a useful
dimensionless parameter that at least approximately characterizes the
operating state of a dendrite.

\subsection{Tip Shape}

The typical scale for dendritic structures is set by the radius of
curvature $\rho$ at the tip.  Very close to the tip, the shape is
approximately a paraboloid of revolution, but for a cubic material
such as NH$_4$Cl, it becomes systematically wider in the planes of the
sidebranches and develops fins.  The sidebranches ultimately develop on
top of those fins.  In the present work, we only consider crystals where
the sidebranches grow in the plane of the image, as in Fig.~\ref{fig:img}
below.

For small four-fold anisotropy, Ben Amar and
Brener \cite{BenAmar1993a} found that the lowest-order correction to
the parabolic shape in the plane containing the sidebranches is
\begin{equation}
\frac{z}{\rho}= \frac{1}{2}\left(\frac{w}{\rho}\right)^2 -
                    A_4 \left(\frac{w}{\rho}\right)^4 ,
\label{eqn:tip4th}
\end{equation}
where the tip is at the origin,
$z$ is the distance from the tip along the axis defined by the main dendrite stem,
$w$ is the half-width of the dendrite,
$\rho$ is the radius of curvature at the tip, and $A_4$ is a small
material-dependent parameter.
This expression is only valid close to the tip, however.  Further back,
Brener \cite{Brener1993} found that the width of the fins deviates
significantly from parabolic and instead scales as a power law
\begin{equation}
\frac{w}{\rho}= a \left(\frac{z}{\rho}\right)^\beta,
\label{eqn:tippower}
\end{equation}
where $a = (5/3)^{3/5}$ and $\beta = 3/5$.
A more complete description of the three-dimensional shape obtained during
phase field simulations is given by Karma, Lee, and Plapp
in Ref.~\cite{Karma2000}.

The net result is that it is challenging to experimentally determine
the tip radius.  Any estimate of $\rho$ involves fitting data over
a finite range of $z$.  At small $z$, data may be limited by the
experimental resolution, while at larger $z$, incipient sidebranches
become more important.  Intermediate $z$ values may span the crossover
from Eq.~\ref{eqn:tip4th} to Eq.~\ref{eqn:tippower}.

In the simulations of Ref.~\cite{Karma2000}, Karma \textit{et al.}\ found that
the fitting parameters for both the fourth-order and power-law
fits depended on the maximum distance $z$ used in the fit, though that
variability also depended on the underlying anisotropy.  For xenon
dendrites, Bisang and Bilgram \cite{Bisang1996} found that the
power-law shape Eq.~\ref{eqn:tippower} provided a more robust fit
to the data than Eq.~\ref{eqn:tip4th}.  On the other hand, Lacombe,
Koss, and Glicksman \cite{Lacombe2007a} found that a hyperbolic
shape worked well for pivalic acid dendrites.

For the growth of ammonium chloride crystals at small supersaturation,
Dougherty and Lahiri \cite{Dougherty2005} determined that using
data up to a distance of $6 \rho$ in Eq.~\ref{eqn:tip4th} gave a
more robust fit than the power law, and found $A_4 = 0.004 \pm
0.001$.  Melendez and Beckermann \cite{Beckermann2012} found a
similar shape correction worked well for succinonitrile-acetone alloys.
LaCombe \textit{et al.}\ also found a similar shape correction
worked well for succinonitrile \cite{LaCombe1995}.
In this work, we follow Ref.~\cite{Dougherty2005} and use Eq.~\ref{eqn:tip4th}
to estimate the tip radius.

A typical crystal image, along with the fit to Eq.~\ref{eqn:tip4th},
is shown in Fig.~\ref{fig:img}.
Two sets of sidebranches are visible in the
plane of the image; two additional sets are growing perpendicular
to the plane along the main stem of the dendrite.

\begin{figure}[htb!]
\includegraphics[width=0.80\columnwidth]{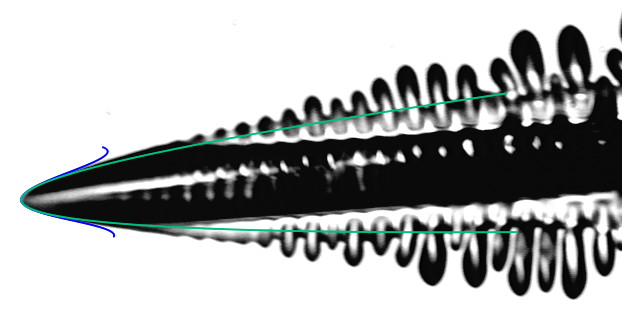}
    \caption{Ammonium chloride dendrite growing from supersaturated
    aqueous solution.  The image is $\SI{390}{\micro m}$ across.
    The tip radius is $\SI{3.1}{\micro m}$ and the growth speed is
    $\SI{1.6}{\micro m/s}$.  The inner curve (green) is the best-fit
    parabola for the tip.  The fourth-order fit (Eq.~\ref{eqn:tip4th})
    is on the far left in blue; the negative fourth-order term causes
    the fit to deviate sharply from the tip before the sidebranches
    become significant.  In this image, the first sidebranches are
    visible approximately $31 \rho$ behind the tip, and have an initial
    wavelength of approximately $4.6 \rho$.}
    \label{fig:img}
\end{figure}

\subsection{Sidebranches}

A short distance behind the tip, sidebranches emerge on top of the fins
with a characteristic
wavelength $\lambda$ that is typically about $3 \sim 6 \rho$.  These
sidebranches start out approximately uniform, but compete in a complex
nonlinear coarsening process \cite{Glicksman1981a, Dougherty1987}.
Larger branches continue to grow, while shorter ones stop, or even begin
to dissolve back, eventually giving rise to structures with a wide
range of length scales.

There have been several approaches to model the origin of sidebranches.
One possibility is that the tip growth itself is actually slightly
oscillatory, and the emerging sidebranches reflect that underlying
oscillation \cite{LaCombe2002, Glicksman2007, Xu2001}.  Though
oscillations were observed in pivalic acid \cite{LaCombe2002}, no such
oscillations were detected in xenon \cite{Bisang1996}, ammonium
bromide \cite{Dougherty1987}, or in the present work.
Another recent line of analysis by Glicksman \cite{Glicksman2015-review,
Glicksman2016,Mullis2015} considers the importance of capillary-mediated interface
perturbations in driving deterministic branching.

The conventional approach is to model sidebranches as arising from
microscopic noise.  The approximate periodicity of
the sidebranches is the result of the selective amplification of that
noise \cite{Langer1987, Dougherty1987, Brener1995, Karma1999, Gonzalez-Cinca2004a,
Wittwer2006}.

Several experiments have shown that applied perturbations can produce
periodic sidebranches.
Directional solidification experiments with pivalic acid/coumarin alloys
have shown that directly applying a pulsing laser to dendritic tips
can drive an oscillating tip and produce a
sidebranch structure with the corresponding wavelength
\cite{Williams1993}.
Similarly, mechanical vibrations have been shown to induce sidebranches
in the free dendritic growth of xenon \cite{Wittwer2006}.
Analogous results have also been obtained in viscous fingering experiments
\cite{Rabaud1988}.
These experiments show that oscillating-tip
solutions are possible, but do not address the question of what happens as
the amplitude of the applied noise is reduced to zero.

For directional solidification, sidebranches have been observed in
bursts that are coherent within a burst, but uncorrelated between
bursts \cite{Georgelin2007}.  This situation has been extensively studied
computationally by Echebarria \textit{et al.}~\cite{Echebarria2010}. They
find that in the presence of a thermal gradient, both noise-induced and
limit cycle branches are possible, which likely explains the coherent
bursts seen in Ref.~\cite{Georgelin2007}.  In the absence of a thermal
gradient, however, as in free dendritic growth, phase field studies
such as those by Karma and Rappel \cite{Karma1999} show that persistent
sidebranches only appear in the presence of noise.

For free dendritic growth, Langer showed how the selective amplification
of noise near the dendrite tip could generate trains of sidebranches
\cite{Langer1987}.  For the axisymmetric case, he found that the resulting
branches were qualitatively similar to those observed in experiments,
but estimated that the predicted amplitude was smaller than observed.
Brener and Temkin \cite{Brener1995} extended that work to non-axisymmetric
growth, where the needle crystal develops fins in the directions favored by
anisotropy, and the sidebranches develop on top of those fins.  This led
to an increased rate of growth.  Specifically, they found that the
noise-induced sidebranch amplitude
$A(z)$ is given by a stretched exponential of the form
\begin{equation}
A(z) = \rho S_0 \exp{ \left[ \frac{2}{3}
    \left( \frac{\bar{w}^3(z)}{3 \sigma^* z \rho^2} \right)^{1/2} \right] } ,
\label{eqn:sbrnoise1}
\end{equation}
where $\bar{w}(z)$ is the average width of the dendrite given by
Eq.~\ref{eqn:tippower}, and $S_0$ is the dimensionless noise amplitude.
For ammonium bromide dendrites growing from supersaturated aqueous
solution, Gonzalez-Cinca \textit{et al.}\ \cite{Gonzalez-Cinca2001}
give $S_0 \sim 6 \times 10^{-5}$ as a conservative over-estimate of the
noise value.  The distance to the first sidebranch, $z_\mathrm{sbr}$,
can then be defined as the distance for which $A(z)$ reaches some arbitrary
fraction of $\rho$.

For the initial sidebranch spacing,
Brener and Temkin \cite{Brener1995} predicted
\begin{equation}
\frac{\lambda(z)}{\rho} = 2 \pi \left(\frac{3}{5}\right)^{3/10}
        \sqrt{3 \sigma^*} \left(\frac{z}{\rho}\right)^{1/5}.
\label{eqn:lambda}
\end{equation}

Beyond the initial development of the sidebranches, a number of
approaches have been developed to model the interaction and coarsening
of sidebranches.  Phase field models attempt to incorporate the full
physics of the three-dimensional problem \cite{Boettinger2002, Chen2002}.
The addition of noise to phase field simulations has also been shown to
produce sidebranching structures similar to those observed in experiments
\cite{Karma1999, Pavlik2000, Tong2001}, though it continues to prove
challenging to perform the calculations in the small undercooling and
small anisotropy range appropriate for ammonium chloride solution
growth \cite{Yamanaka2011, Lin2011}.
A variety of numerical models and approaches are reviewed in
Jaafar \textit{et al.}~\cite{Jaafar2017}.

Experimentally, power law behavior has also been reported for a number
of integral parameters of dendritic growth, including the sidebranch
envelope, contour area, and volume \cite{Beckermann1998, Hurlimann1992,
Bilgram1993a, Bisang1996, Singer2006, Wittwer2006}.  In contrast, the
average width of both pivalic acid and ammonium chloride dendrites was
found to follow a simple power law only over a limited range of $z$
values \cite{Dougherty1994}.  These results all highlight the complex ways
sidebranches interact beyond their initial development.

\section{Materials and Methods}


The experiments were performed with a solution of ammonium chloride
(Fisher Scientific, 99.99\%) in water (Fisher Scientific, HPLC grade,
filtered through a $\SI{0.1}{\micro \meter}$ filter).
The concentration was approximately 36\% NH$_4$Cl by weight,
for a saturation temperature of approximately \SI{66.7}{\celsius}.
The solution
was placed in a $40 \times 10 \times 2$~mm$^3$ glass spectrophotometer cell
sealed with a Teflon stopper held in place by epoxy.
The cell was mounted in a temperature-controlled copper block, surrounded by a
temperature-controlled outer aluminum block,
and placed on an insulated microscope stage.  The entire apparatus was
enclosed in a temperature-controlled insulated plexiglas box.
The rms temperature fluctuations of the copper block
were approximately $\SI{2e-4}{\celsius}$.
Temperature gradients were estimated to be less than
$\SI{0.001}{\celsius / mm}$.
Additional details of the
experimental apparatus and protocol are given in Ref.~\cite{Dougherty2019a}.

The solution was heated to dissolve all the solids, stirred to eliminate
concentration gradients, and then cooled.
Upon cooling, many crystals would nucleate.  The system was then warmed
until only one seed remained.  That seed was held in equilibrium and
then cooled at a rate of $\SI{-6.0e-4}{\celsius/s}$ to allow a single
dendrite to develop and grow.  Once that crystal became well-established,
the cooling rate was increased to
$\SI{-1.6e-3}{\celsius/s}$.  Because the finite cell became depleted as
the crystal grew, it was necessary to continually lower the temperature
to maintain growth throughout the $\SI{10000}{s}$ run.  Even at the given
cooling rate, however, the crystal did slow significantly; overall the
tip velocity varied from $\SI{7.2}{\micro m/s}$ to $\SI{0.8}{\micro m/s}$.
The change in velocity was generally quite slow, however.  Over the typical
distance and time scales corresponding to the initial growth of sidebranches,
the velocity changes were between 1\% and 4\%.

Although the temperature varied slowly during the course of a run, we still
model the growth as that of an isothermal mixture limited by chemical diffusion.
For ammonium chloride solutions \cite{AmmoniumChloride2018}, the thermal
diffusion constant is approximately $\alpha = \SI{1.3e5}{\micro m^2/s}$, while
we estimate \cite{Dougherty2019a} the chemical diffusion constant to be
approximately $D = \SI{2500}{\micro m^2/s}$.  The Lewis number is then
$\alpha / D \approx 50$, indicating that the process is dominated by chemical
diffusion.

\subsection{Imaging}

Images were obtained at 1-second intervals from a charged coupled device
(CCD) camera attached to the microscope and acquired directly into the
computer via a Data Translation DT3155 frame grabber with a resolution
of $640 \times 480$ pixels.  The resolution of the images was
$\SI{0.628 \pm 0.010}{\micro \meter/pixel}$.  A typical crystal image,
with $\rho = \SI{3.1}{\micro m}$ and
$v = \SI{1.6}{\micro m/s}$,
is shown above in Fig.~\ref{fig:img}.

The interface position was determined by an iterative process, described
in detail in Ref.~\cite{Dougherty2005}.
Briefly, the image intensity was scanned on lines roughly perpendicular
to the interface.  Over the range of about 4 pixels, the intensity
dropped rapidly from the outside to the inside of the crystal.  In that
transition region, a straight line was fit to the intensity function,
and the border was interpolated as the position where that fit intensity
equaled the average of the intensity just outside and just inside the
crystal.  Those border points were used to make initial estimates
of the tip position, orientation, $\rho$ and $A_4$.
Those initial estimates were
then used to run a new set of image scans perpendicular to the interface,
and the process was iterated until it converged.
A final set of scans was run along the full length of the dendrite
to measure the dendrite width $w(z)$.

To fit to Eq.~\ref{eqn:tip4th}, only data with $z \le z_\mathrm{max} =
6 \rho$ was used. As was shown in Ref.~\cite{Dougherty2005}, this is a
compromise value.  Although Eq.~\ref{eqn:tip4th} applies close to the
tip, the small and highly-curved tip is the most difficult part to image
accurately, so fits with a small $z_\mathrm{max}$ tend to be less robust.
On the other hand, fits with larger $z_\mathrm{max}$ may start to include
early sidebranches, and also enter a regime where Eq.~\ref{eqn:tip4th}
is no longer appropriate.  (See also Ref.~\cite{Beckermann2012} for a
similar discussion for succinonitrile-acetone alloys.)

\section{Results}

\begin{figure}[htb!]
\centering
\includegraphics[width=0.90\columnwidth]{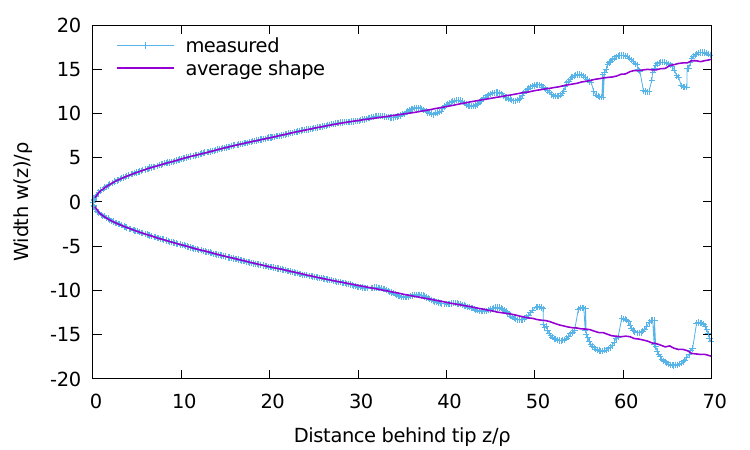}
    \caption{Width of a portion of the crystal in Fig.~\ref{fig:img}
    as a function of distance back from the tip.  All distances are scaled
    by the tip radius $\rho$, and only the first few sidebranches are shown.
    The solid line is the measured scaled width averaged over all crystals
    grown under similar conditions.}
    \label{fig:width}
\end{figure}

The width of the early sidebranch region
of the crystal from Fig.~\ref{fig:img} is shown in
Fig.~\ref{fig:width},
along with the measured average shape $\bar{w}(z)$ for all crystals
grown under similar conditions.  All distances have been scaled by the
tip radius $\rho$.

\subsection{Tip Radius and Velocity}

\begin{figure}[htb!]
\centering
\includegraphics[width=0.90\columnwidth]{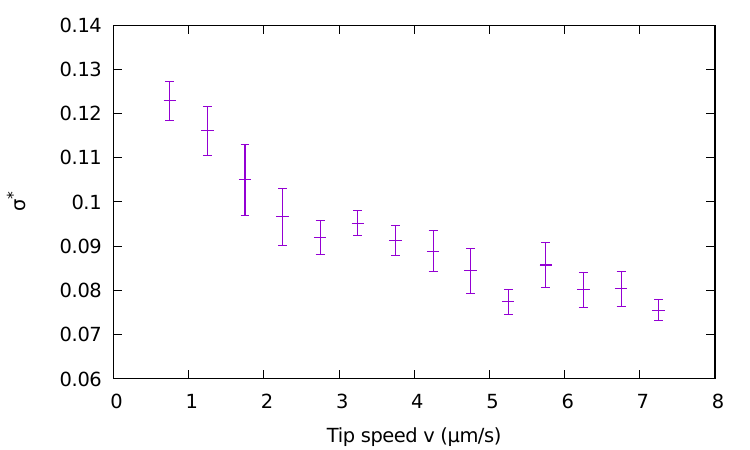}
    \caption{Variation of the parameter $\sigma^*$ with
    growth speed.
    The error bars indicate one standard deviation.
    There is a small downward trend with velocity.
    }
    \label{fig:sigmastar}
\end{figure}

As the cell became depleted over the course of the run, the tip velocity
gradually changed from $\SI{7.2}{\micro m/s}$ to $\SI{0.8}{\micro m/s}$,
while the tip radius varied from $\SI{1.5}{\micro m}$ to $\SI{4.0}{\micro
m}$.  The value of the combination $D d_0$ for this material was
previously reported \cite{Dougherty2019a} to be $D d_0 = \SI{0.78
\pm 0.07}{\micro m^3/s}$.  The resulting values for $\sigma^*$
in Eq.~\ref{eqn:sigmastar} are shown in Fig.~\ref{fig:sigmastar}.
There is a downward trend with velocity, as was seen in previous experiments
\cite{Glicksman1995a, Lacombe2007a, Beckermann2012, Martyushev2016}
as well as phase field simulations
\cite{Mullis2011, Bollada2015}.
For this work, we used the average value $\sigma ^* = 0.10 \pm 0.02$.

\subsection{Sidebranches}

In Fig.~\ref{fig:width}, the first sidebranches are visible starting around
$z \approx 31 \rho$, and the initial spacing is $\lambda = 4.6 \rho$.
Beyond $z \sim 50 \rho$, significant competition between sidebranches
clearly affects their growth.

We first measured the envelope of active
sidebranches, as in Refs.~\cite{Wittwer2006} and \cite{Beckermann2012}.
A sidebranch was considered ``active'' if it was larger than all other
sidebranches on the same side closer to the tip.  A sidebranch also had
to be at least a distance $\rho$ away from the previous branch in order
to be considered a new branch.

\begin{figure}[htb!]
\centering
\includegraphics[width=0.90\columnwidth]{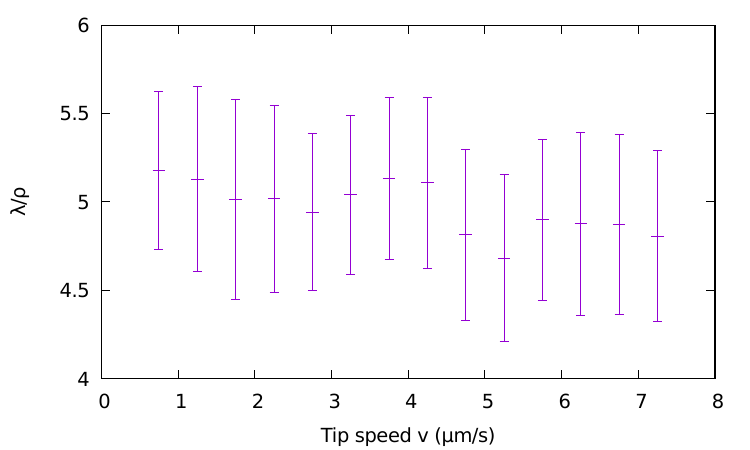}
    \caption{Mean sidebranch wavelength as a function of growth speed.
    Error bars indicate one standard deviation.
    There is a very slight downward trend with velocity, but the typical
    variations at any speed are larger than the overall trend.
    }
    \label{fig:lambda}
\end{figure}

The average sidebranch wavelength $\lambda$ was estimated by performing
a linear fit to the positions of adjacent early sidebranches \textit{vs.}\
sidebranch number.  The results are shown in Fig.~\ref{fig:lambda}.
There was considerable variation in spacing from image to image, but
the overall mean value was $\lambda = (5.05 \pm 0.01)\rho$, where the
uncertainty is the standard deviation of the mean.

\begin{figure}[htb!]
\centering
\includegraphics[width=0.90\columnwidth]{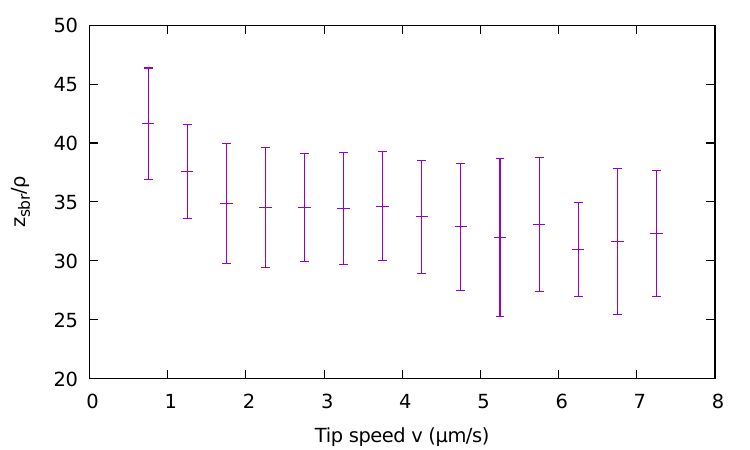}
    \caption{Position of first sidebranch as a function of growth speed.
    Error bars indicate one standard deviation.
    There is a slight downward trend with velocity, although the typical
    variations at any speed are comparable to the overall trend.
    }
    \label{fig:zsbr}
\end{figure}

We also estimated the distance $z_\mathrm{sbr}$ to the first detectable
sidebranch.  Since the identification of the first branch tended to be
significantly affected by noise, the following procedure was adopted:
The position and amplitude of the first 4 branches
were found and fit to a straight line.  The amplitude was found
by measuring the deviation of the width $w(z)$ from the measured average
shape $\bar{w}(z)$.  The distance $z_\mathrm{sbr}$ was taken to be the
distance at which the fit sidebranch amplitude would equal an arbitrary
threshold of $0.25 \rho$.
For the crystal in Fig.~\ref{fig:width}, this gives $z_\mathrm{sbr}
= 32.8$, but there was considerable variation throughout the run.
Results for all crystals are shown in Fig.~\ref{fig:zsbr}.
Overall, the average distance was
$z_\mathrm{sbr} = (35.8 \pm 0.1) \rho$, where the
uncertainty is one standard deviation of the mean.

Using this value for $z$ in Eq.~\ref{eqn:lambda}, we would expect
$\lambda(z_\mathrm{sbr}) = (6.1 \pm 0.3) \rho$, somewhat larger than
the measured value of $5.05 \pm 0.01$.  Conversely, constraining
$\lambda$ to the measured value in Eq.~\ref{eqn:lambda} would require
$z_\mathrm{sbr} \sim 14\rho$.
(Since Eq.~\ref{eqn:lambda} depends only weakly on $z$,
the dominant uncertainty is that in $\sigma^*$.)

Finally, we considered the amplitude of early sidebranches, and compared
with the theoretical noise prediction from Eq.~\ref{eqn:sbrnoise1}.
Since most of the sidebranch parameters vary only slowly with growth speed,
we scaled all crystals by the appropriate tip radius $\rho$.

To estimate $A(z)$, we
computed the root mean square (rms) deviation of the measured shape
around $\bar{w}(z)$.
For $\bar{w}(z)$, we considered several models, including
the power law of Eq.~\ref{eqn:tippower}, a parabola $\sqrt{2 \rho z}$
with higher-order corrections, and a hyperboloid, but none
fit the data sufficiently well over the range of interest.  Instead,
we used the actual measured average shape.

\begin{figure}[htb!]
\centering
\includegraphics[width=0.90\columnwidth]{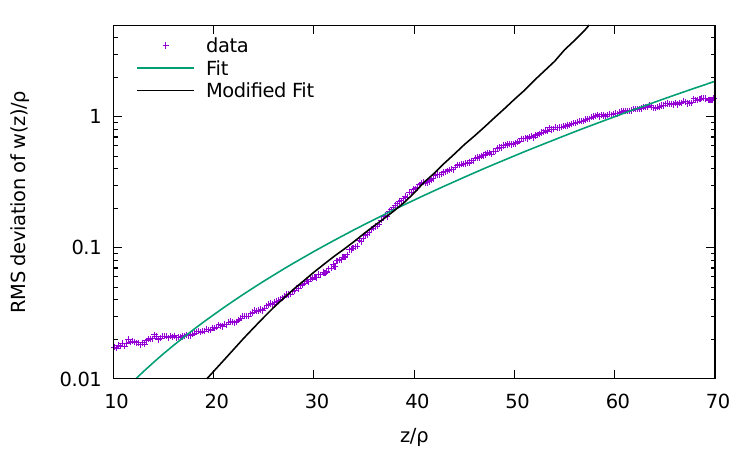}
    \caption{Semi-log plot of the measured sidebranch amplitude as a function
    of distance from the tip, along with the best fit to Eq.~\ref{eqn:sbrnoise1},
    and a fit to the modified prediction in Eq.~\ref{eqn:sbrnoise2}.
    Beyond about $z = 40\rho$, competition among the sidebranches becomes more
    important and Eq.~\ref{eqn:sbrnoise2} no longer applies.
    }
    \label{fig:sbramp}
\end{figure}

The results are shown in Fig.~\ref{fig:sbramp}, along with
a fit to the prediction in Eq.~\ref{eqn:sbrnoise1} from Ref.~\cite{Brener1995}, with a noise
amplitude of $S_0 = (5.6 \pm 0.1) \times 10^{-5}$.
It is clear that the amplitude grows more rapidly than the model in the
early sidebranch regime, so
we also considered a modified fit, where we included
an additional dimensionless term $s$ in the exponential factor:
\begin{equation}
A(z) = S_0 \exp{ \left[ s \frac{2}{3}
    \left( \frac{\bar{w}^3(z)}{3 \sigma^* z} \right)^{1/2} \right] } ,
\label{eqn:sbrnoise2}
\end{equation}
where all distances have been scaled by $\rho$.
The best-fit parameters are $S_0 = (1.1 \pm 0.2) \times 10^{-6}$ and
$s = 1.72 \pm 0.03$.  The value for $S_0$ is less than the conservative
over-estimate of $S_0 \sim 6 \times 10^{-5}$ given by Gonzalez-Cinca
\textit{et al.}\ \cite{Gonzalez-Cinca2001}, but the value for $s$ is
not in agreement with the expected value of 1 in Eq.~\ref{eqn:sbrnoise1}.

It is worth emphasizing that the fit is only applicable over a rather
narrow range of $z$ values.  At small distances, $z \lesssim 20 \rho$,
the measurements are dominated by noise, while at larger distances, $z
\gtrsim 40 \rho$, competition among sidebranches slows down their growth.
Similar constraints were also apparent in the study of xenon sidebranches
\cite{Wittwer2006}, where the theoretical fit worked well only in the
range of $8 \sim 15 \rho$.

\begin{figure}[htb!]
\centering
\includegraphics[width=0.90\columnwidth]{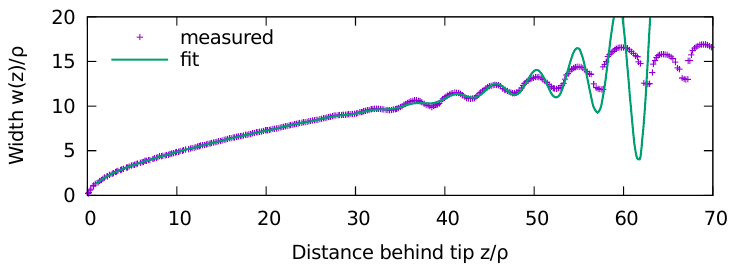}
    \caption{Width of one side of the crystal from Fig.~\ref{fig:width}
    along with a fit to Eq.~\ref{eqn:indiv}.  At larger distances,
    as in Fig.~\ref{fig:sbramp}, the sidebranches enter a
    nonlinear competitive regime where the fit is not applicable.
    }
    \label{fig:indiv}
\end{figure}

As an additional check, we extended
Eq.~\ref{eqn:sbrnoise2} to include an oscillatory term so that it could
be applied to individual images, such as Fig.~\ref{fig:width}:
\begin{equation}
w(z) = \bar{w}(z) + A(z) \sin\left(\frac{2 \pi}{\lambda} + \phi \right) ,
\label{eqn:indiv}
\end{equation}
where $\lambda$ is the wavelength for that particular image and $\phi$
is the phase.  The best fit, with
$\lambda = 4.6$, is shown in Fig.~\ref{fig:indiv}.  The results from these
individual fits were consistent with the global average fit above.

\section{Discussion}

We have considered the early stages of sidebranching in the growth
of ammonium chloride dendrites from supersaturated aqueous solution.
When all distances are scaled by the tip radius,
the overall features vary only slightly with growth speed, and are
roughly consistent with what would be expected if the sidebranches are
due to the selective amplification of noise.

A stretched exponential of the form of Eq.~\ref{eqn:sbrnoise2} does
provide a reasonable fit to the sidebranch amplitude data, but
that fit requires an additional factor $s$ in the exponential.
The noise amplitude $S_0$ is consistent with that estimated for intrinsic
noise \cite{Gonzalez-Cinca2001}, and the resulting distance to the
first sidebranch, $z_\mathrm{sbr}$, is also consistent with the theory.

By contrast for xenon, Wittwer and Bilgram \cite{Wittwer2006} found good
agreement for Eq.~\ref{eqn:sbrnoise1} without any additional factor $s$.
The reason for the difference is not clear, but it is worth noting that
the value for $\sigma^* = 0.10 \pm 0.02$ found in the present work is
five times the value $\sigma^* = 0.02$ used in Ref.~\cite{Wittwer2006},
and the distance to the first sidebranch $z_\mathrm{sbr}$ in ammonium
chloride is roughly twice the value $z_\mathrm{sbr}/\rho = 17.5 \pm 3$
reported for xenon \cite{Wittwer2006}.
Whether these discrepancies are due to different material properties,
solution \textit{vs.}\ thermal growth, or some other factor, is unclear.

The measured mean sidebranch wavelength $\lambda = (5.05 \pm 0.01)\rho$
for ammonium chloride is slightly smaller than the predicted value
of $(6.1 \pm 0.3) \rho$ from Eq.~\ref{eqn:lambda}.  For succinonitrile
\cite{Glicksman1981a, Langer1987, Brener1995, Corrigan1999a}, the measured
value was consistent with the theory, while for pivalic acid, the measured
value was slightly larger than predicted \cite{Giummarra2005a}.  The
source of these small differences remains unclear.

It is important to recognize that the range of applicability of
Eq.~\ref{eqn:sbrnoise2} is quite limited, so the fitted function is not
particularly well-constrained.  At small $z$, there are two main issues.
First, emerging sidebranches are potentially masked by measurement noise.
Second, measurements of the tip itself (and hence all distances scaled
by $\rho$) are also potentially contaminated by early sidebranches.
These effects are particularly problematic near the tip because the
concentration gradients and corresponding optical distortions are largest
there.  Measurements of the very early sidebranches and tip shape are
thus inextricably intertwined.  At larger $z$, the increasing nonlinear
competition among sidebranches leads to changes in the expected scaling
behavior.  At even larger $z$, it is no longer possible to characterize
the structure by a single-valued width function $w(z)$.  These constraints
make it much more challenging to identify whether there are simple
underlying scaling laws that govern the origin and initial growth of sidebranches.

\section*{Acknowledgments}
The author thanks Lafayette College for financial support.
The author acknowledges help from former Lafayette College students
Thomas Nunnally and Raluca Eftimoiu
with preliminary versions of these experiments.
The author also thanks an anonymous referee for helpful comments.

This research did not receive any specific grant from funding agencies
in the public, commercial, or not-for-profit sectors.

Data used in this paper is available through Mendeley at
\url{http://dx.doi.org/10.17632/sdxdsm6trn.1}.

Declarations of interest: none.



\bibliographystyle{elsarticle-num}
\bibliography{sbramp}

\begin{thebibliography}{10}
\expandafter\ifx\csname url\endcsname\relax
  \def\url#1{\texttt{#1}}\fi
\expandafter\ifx\csname urlprefix\endcsname\relax\def\urlprefix{URL }\fi
\expandafter\ifx\csname href\endcsname\relax
  \def\href#1#2{#2} \def\path#1{#1}\fi

\bibitem{Langer1980}
J.~S. Langer, Instabilities and pattern formation in crystal growth, Rev. Mod.
  Phys. 52~(1) (1980) 1--28.
\newblock \href {https://doi.org/10.1103/RevModPhys.52.1}
  {\path{doi:10.1103/RevModPhys.52.1}}.

\bibitem{Glicksman2015-review}
M.~E. Glicksman, 16 - {{Dendritic Growth}}, in: T.~Nishinaga (Ed.), Handbook of
  {{Crystal Growth}} ({{Second Edition}}), {Elsevier}, Boston, 2015, pp.
  669--722.
\newblock \href {https://doi.org/10.1016/B978-0-444-56369-9.00016-2}
  {\path{doi:10.1016/B978-0-444-56369-9.00016-2}}.

\bibitem{Jaafar2017}
M.~A. Jaafar, D.~R. Rousse, S.~Gibout, J.-P. Bedecarrats, A review of dendritic
  growth during solidification: {{Mathematical}} modeling and numerical
  simulations, Renew. Sust. Energ. Rev. 74 (2017) 1064--1079.
\newblock \href {https://doi.org/10.1016/j.rser.2017.02.050}
  {\path{doi:10.1016/j.rser.2017.02.050}}.

\bibitem{Asta2009}
M.~Asta, C.~Beckermann, A.~Karma, W.~Kurz, R.~Napolitano, M.~Plapp, G.~Purdy,
  M.~Rappaz, R.~Trivedi, Solidification microstructures and solid-state
  parallels: Recent developments, future directions, Acta Mater. 57~(4) (2009)
  941--971.
\newblock \href {https://doi.org/10.1016/j.actamat.2008.10.020}
  {\path{doi:10.1016/j.actamat.2008.10.020}}.

\bibitem{Trivedi2000}
W.~J. Boettinger, S.~R. Coriell, A.~L. Greer, A.~Karma, W.~Kurz, M.~Rappaz,
  R.~Trivedi, Solidification microstructures: {{Recent}} developments, future
  directions, Acta Mater. 48~(1) (2000) 43--70.
\newblock \href {https://doi.org/10.1016/S1359-6454(99)00287-6}
  {\path{doi:10.1016/S1359-6454(99)00287-6}}.

\bibitem{Akamatsu2016}
S.~Akamatsu, H.~{Nguyen-Thi}, In situ observation of solidification patterns in
  diffusive conditions, Acta Mater. 108 (2016) 325--346.
\newblock \href {https://doi.org/10.1016/j.actamat.2016.01.024}
  {\path{doi:10.1016/j.actamat.2016.01.024}}.

\bibitem{Huang2012}
W.~Huang, L.~Wang, Solidification researches using transparent model materials
  \textemdash{} {{A}} review, Science China Technological Sciences 55~(2)
  (2012) 377--386.
\newblock \href {https://doi.org/10.1007/s11431-011-4689-1}
  {\path{doi:10.1007/s11431-011-4689-1}}.

\bibitem{Rolley1994}
E.~Rolley, S.~Balibar, F.~Graner, Growth shape of $^{3}\mathrm{He}$ needle
  crystals, Phys. Rev. E 49 (1994) 1500--1506.
\newblock \href {https://doi.org/10.1103/PhysRevE.49.1500}
  {\path{doi:10.1103/PhysRevE.49.1500}}.

\bibitem{Bilgram1993}
J.~Bilgram, E.~Hurlimann, Dendritic {{Solidification}} of {{Rare}}-{{Gases}},
  Prog. Cryst. Growth Charact. Mater. 26 (1993) 67--86.
\newblock \href {https://doi.org/10.1016/0960-8974(93)90010-2}
  {\path{doi:10.1016/0960-8974(93)90010-2}}.

\bibitem{Dougherty1987}
A.~Dougherty, P.~Kaplan, J.~Gollub, Development of {{Side Branching}} in
  {{Dendritic Crystal}}-{{Growth}}, Phys. Rev. Lett. 58~(16) (1987) 1652--1655.
\newblock \href {https://doi.org/10.1103/PhysRevLett.58.1652}
  {\path{doi:10.1103/PhysRevLett.58.1652}}.

\bibitem{Couder2005}
Y.~Couder, J.~Maurer, R.~{Gonzalez-Cinca}, A.~{Hernandez-Machado}, Side-branch
  growth in two-dimensional dendrites. {{I}}. {{Experiments}}, Phys. Rev. E
  71~(3) (2005) 031602--031602.
\newblock \href {https://doi.org/10.1103/PhysRevE.71.031602}
  {\path{doi:10.1103/PhysRevE.71.031602}}.

\bibitem{Honjo1985}
H.~Honjo, S.~Ohta, Y.~Sawada, New {{Experimental Findings}} in
  {{Two}}-{{Dimensional Dendritic Crystal}}-{{Growth}}, Phys. Rev. Lett. 55~(8)
  (1985) 841--844.
\newblock \href {https://doi.org/10.1103/PhysRevLett.55.841}
  {\path{doi:10.1103/PhysRevLett.55.841}}.

\bibitem{Tanaka1992}
A.~Tanaka, M.~Sano, Measurement of the kinetic effect on the concentration
  field of a growing dendrite, J. Cryst. Growth 125~(1–2) (1992) 59--64.
\newblock \href {https://doi.org/10.1016/0022-0248(92)90320-I}
  {\path{doi:10.1016/0022-0248(92)90320-I}}.

\bibitem{Sawada1995}
T.~Sawada, K.~Takemura, K.~Shigematsu, S.~ichi Yoda, K.~Kawasaki, Diffusion
  field around a dendrite growing under microgravity, Phys. Rev. E 51 (1995)
  R3834--R3837.
\newblock \href {https://doi.org/10.1103/PhysRevE.51.R3834}
  {\path{doi:10.1103/PhysRevE.51.R3834}}.

\bibitem{Dougherty2005}
A.~Dougherty, M.~Lahiri, Shape of ammonium chloride dendrite tips at small
  supersaturation, J. Cryst. Growth 274~(1-2) (2005) 233--240.
\newblock \href {https://doi.org/10.1016/j.jcrysgro.2004.09.065}
  {\path{doi:10.1016/j.jcrysgro.2004.09.065}}.

\bibitem{Martyushev2016}
L.~M. Martyushev, P.~S. Terentiev, A.~S. Soboleva, Unsteady growth of ammonium
  chloride dendrites, J. Cryst. Growth 436 (2016) 82--86.
\newblock \href {https://doi.org/10.1016/j.jcrysgro.2015.12.002}
  {\path{doi:10.1016/j.jcrysgro.2015.12.002}}.

\bibitem{Glicksman1981a}
S.~C. Huang, M.~E. Glicksman, Overview 12: {{Fundamentals}} of dendritic
  solidification\textemdash{{II}} development of sidebranch structure, Acta
  Metall. 29~(5) (1981) 717--734.
\newblock \href {https://doi.org/10.1016/0001-6160(81)90116-4}
  {\path{doi:10.1016/0001-6160(81)90116-4}}.

\bibitem{Corrigan1999a}
D.~P. Corrigan, M.~B. Koss, J.~C. LaCombe, K.~D. {de Jager}, L.~A. Tennenhouse,
  M.~E. Glicksman, Experimental measurements of sidebranching in thermal
  dendrites under terrestrial-gravity and microgravity conditions, Phys. Rev. E
  60~(6) (1999) 7217--7223.
\newblock \href {https://doi.org/10.1103/PhysRevE.60.7217}
  {\path{doi:10.1103/PhysRevE.60.7217}}.

\bibitem{Beckermann1998}
Q.~Li, C.~Beckermann, Scaling behavior of three-dimensional dendrites, Phys.
  Rev. E 57~(3) (1998) 3176--3188.
\newblock \href {https://doi.org/10.1103/PhysRevE.57.3176}
  {\path{doi:10.1103/PhysRevE.57.3176}}.

\bibitem{Giummarra2005a}
C.~Giummarra, J.~C. LaCombe, M.~B. Koss, J.~E. Frei, A.~O. Lupulescu, M.~E.
  Glicksman, Sidebranch characteristics of pivalic acid dendrites grown under
  convection-free and diffuso-convective conditions, J. Cryst. Growth 274~(1)
  (2005) 317--330.
\newblock \href {https://doi.org/10.1016/j.jcrysgro.2004.10.039}
  {\path{doi:10.1016/j.jcrysgro.2004.10.039}}.

\bibitem{Dougherty1994}
A.~Dougherty, A.~Gunawardana, Mean {{Shape}} of 3-{{Dimensional Dendrites}} - a
  {{Comparison}} of {{Pivalic Acid}} and {{Ammonium}}-{{Chloride}}, Phys. Rev.
  E 50~(2) (1994) 1349--1352.
\newblock \href {https://doi.org/10.1103/PhysRevE.50.1349}
  {\path{doi:10.1103/PhysRevE.50.1349}}.

\bibitem{Bisang1996}
U.~Bisang, J.~H. Bilgram, Shape of the tip and the formation of sidebranches of
  xenon dendrites, Phys. Rev. E 54~(5) (1996) 5309--5326.
\newblock \href {https://doi.org/10.1103/PhysRevE.54.5309}
  {\path{doi:10.1103/PhysRevE.54.5309}}.

\bibitem{Wittwer2006}
O.~Wittwer, J.~H. Bilgram, {Three-dimensional xenon dendrites: Characterization
  of sidebranch growth}, {Phys. Rev. E} {74}~({4, Part 1}) (2006) {041604}.
\newblock \href {https://doi.org/10.1103/PhysRevE.74.041604}
  {\path{doi:10.1103/PhysRevE.74.041604}}.

\bibitem{Dougherty1992}
A.~Dougherty, R.~Chen, Coarsening and the {{Mean Shape}} of 3-{{Dimensional
  Dendritic Crystals}}, Phys. Rev. A 46~(8) (1992) R4508--R4511.
\newblock \href {https://doi.org/10.1103/PhysRevA.46.R4508}
  {\path{doi:10.1103/PhysRevA.46.R4508}}.

\bibitem{Kishinawa2008}
K.~Kishinawa, H.~Honjo, H.~Sakaguchi, Scale-invariant competitive growth of
  side branches in a dendritic crystal, Phys. Rev. E 77 (2008) 030602.
\newblock \href {https://doi.org/10.1103/PhysRevE.77.030602}
  {\path{doi:10.1103/PhysRevE.77.030602}}.

\bibitem{Beckermann2012}
A.~J. Melendez, C.~Beckermann, Measurements of dendrite tip growth and
  sidebranching in succinonitrile-acetone alloys, J. Cryst. Growth 340~(1)
  (2012) 175--189.
\newblock \href {https://doi.org/10.1016/j.jcrysgro.2011.12.010}
  {\path{doi:10.1016/j.jcrysgro.2011.12.010}}.

\bibitem{Langer1987}
J.~S. Langer, Dendritic sidebranching in the three-dimensional symmetric model
  in the presence of noise, Phys. Rev. A 36~(7) (1987) 3350--3358.
\newblock \href {https://doi.org/10.1103/PhysRevA.36.3350}
  {\path{doi:10.1103/PhysRevA.36.3350}}.

\bibitem{Brener1995}
E.~Brener, D.~Temkin, Noise-{{Induced Sidebranching}} in the 3-{{Dimensional
  Nonaxisymmetric Dendritic Growth}}, Phys. Rev. E 51~(1) (1995) 351--359.
\newblock \href {https://doi.org/10.1103/PhysRevE.51.351}
  {\path{doi:10.1103/PhysRevE.51.351}}.

\bibitem{Kessler1988}
D.~A. Kessler, J.~Koplik, H.~Levine, Pattern {{Selection}} in {{Fingered Growth
  Phenomena}}, Adv. Phys. 37~(3) (1988) 255--339.
\newblock \href {https://doi.org/10.1080/00018738800101379}
  {\path{doi:10.1080/00018738800101379}}.

\bibitem{Glicksman1995a}
M.~E. Glicksman, M.~B. Koss, L.~T. Bushnell, J.~C. Lacombe, E.~A. Winsa,
  Dendritic {{Growth}} of {{Succinonitrile}} in {{Terrestrial}} and
  {{Microgravity Conditions}} as a {{Test}} of {{Theory}}., ISIJ International
  35~(6) (1995) 604--610.
\newblock \href {https://doi.org/10.2355/isijinternational.35.604}
  {\path{doi:10.2355/isijinternational.35.604}}.

\bibitem{Lacombe2007a}
J.~C. LaCombe, M.~B. Koss, M.~E. Glicksman, Tip {{Velocities}} and {{Radii}} of
  {{Curvature}} of {{Pivalic Acid Dendrites}} under {{Convection}}-{{Free
  Conditions}}, Metall and Mat Trans A 38~(1) (2007) 116--126.
\newblock \href {https://doi.org/10.1007/s11661-006-9018-0}
  {\path{doi:10.1007/s11661-006-9018-0}}.

\bibitem{Mullis2011}
A.~M. Mullis, Prediction of the operating point of dendrites growing under
  coupled thermosolutal control at high growth velocity, Phys. Rev. E 83~(6)
  (2011) 061601.
\newblock \href {https://doi.org/10.1103/PhysRevE.83.061601}
  {\path{doi:10.1103/PhysRevE.83.061601}}.

\bibitem{Bollada2015}
P.~C. Bollada, C.~E. Goodyer, P.~K. Jimack, A.~M. Mullis, Simulations of
  three-dimensional dendritic growth using a coupled thermo-solutal phase-field
  model, Appl. Phys. Lett. 107~(5) (2015) 053108.
\newblock \href {https://doi.org/10.1063/1.4928487}
  {\path{doi:10.1063/1.4928487}}.

\bibitem{BenAmar1993a}
M.~B. Amar, E.~Brener, Theory of {{Pattern Selection}} in 3-{{Dimensional
  Nonaxisymmetric Dendritic Growth}}, Phys. Rev. Lett. 71~(4) (1993) 589--592.
\newblock \href {https://doi.org/10.1103/PhysRevLett.71.589}
  {\path{doi:10.1103/PhysRevLett.71.589}}.

\bibitem{Brener1993}
E.~Brener, Needle-crystal solution in three-dimensional dendritic growth, Phys.
  Rev. Lett. 71~(22) (1993) 3653--3656.
\newblock \href {https://doi.org/10.1103/PhysRevLett.71.3653}
  {\path{doi:10.1103/PhysRevLett.71.3653}}.

\bibitem{Karma2000}
A.~Karma, Y.~Lee, M.~Plapp, Three-dimensional dendrite-tip morphology at low
  undercooling, Physical Review E 61~(4) (2000) 3996--4006.
\newblock \href {https://doi.org/10.1103/PhysRevE.61.3996}
  {\path{doi:10.1103/PhysRevE.61.3996}}.

\bibitem{LaCombe1995}
J.~C. LaCombe, M.~B. Koss, V.~E. Fradkov, M.~E. Glicksman, Three-dimensional
  dendrite-tip morphology, Phys. Rev. E 52~(3) (1995) 2778--2786.
\newblock \href {https://doi.org/10.1103/PhysRevE.52.2778}
  {\path{doi:10.1103/PhysRevE.52.2778}}.

\bibitem{LaCombe2002}
J.~C. LaCombe, M.~B. Koss, J.~E. Frei, C.~Giummarra, A.~O. Lupulescu, M.~E.
  Glicksman, Evidence for tip velocity oscillations in dendritic
  solidification, Phys. Rev. E 65 (2002) 031604.
\newblock \href {https://doi.org/10.1103/PhysRevE.65.031604}
  {\path{doi:10.1103/PhysRevE.65.031604}}.

\bibitem{Glicksman2007}
M.~E. Glicksman, J.~S. Lowengrub, S.~Li, X.~Li, A deterministic mechanism for
  dendritic solidification kinetics, JOM 59~(8) (2007) 27--34.
\newblock \href {https://doi.org/10.1007/s11837-007-0100-x}
  {\path{doi:10.1007/s11837-007-0100-x}}.

\bibitem{Xu2001}
J.-J. Xu, D.-S. Yu, Further examinations of dendritic growth theories, J.
  Cryst. Growth 222~(1) (2001) 399--413.
\newblock \href {https://doi.org/10.1016/S0022-0248(00)00920-9}
  {\path{doi:10.1016/S0022-0248(00)00920-9}}.

\bibitem{Glicksman2016}
M.~E. Glicksman, Capillary-mediated interface perturbations: {{Deterministic}}
  pattern formation, J. Cryst. Growth 450 (2016) 119--139.
\newblock \href {https://doi.org/10.1016/j.jcrysgro.2016.03.031}
  {\path{doi:10.1016/j.jcrysgro.2016.03.031}}.

\bibitem{Mullis2015}
A.~M. Mullis, Spontaneous deterministic side-branching behavior in phase-field
  simulations of equiaxed dendritic growth, J. Appl. Phys. 117~(11) (2015)
  114305.
\newblock \href {https://doi.org/10.1063/1.4915278}
  {\path{doi:10.1063/1.4915278}}.

\bibitem{Karma1999}
A.~Karma, W.-J. Rappel, Phase-field model of dendritic sidebranching with
  thermal noise, Phys. Rev. E 60~(4) (1999) 3614--3625.
\newblock \href {https://doi.org/10.1103/PhysRevE.60.3614}
  {\path{doi:10.1103/PhysRevE.60.3614}}.

\bibitem{Gonzalez-Cinca2004a}
R.~{Gonz\'alez-Cinca}, Y.~Couder, J.~Maurer, A.~{Herna\'andez-Machado}, The
  {{Role}} of {{Noise}} in {{Sidebranching Development}}, Fluctuation \& Noise
  Letters 4~(4) (2004) L535--L544.
\newblock \href {https://doi.org/10.1142/S0219477504002178}
  {\path{doi:10.1142/S0219477504002178}}.

\bibitem{Williams1993}
L.~M. Williams, M.~Muschol, X.~Qian, W.~Losert, H.~Z. Cummins, Dendritic
  sidebranching with periodic localized perturbations: {{Directional}}
  solidification of pivalic acid--coumarin 152 mixtures, Phys. Rev. E 48~(1)
  (1993) 489--499.
\newblock \href {https://doi.org/10.1103/PhysRevE.48.489}
  {\path{doi:10.1103/PhysRevE.48.489}}.

\bibitem{Rabaud1988}
M.~Rabaud, Y.~Couder, N.~Gerard, Dynamics and stability of anomalous
  {{Saffman}}-{{Taylor}} fingers, Phys. Rev. A 37~(3) (1988) 935--947.
\newblock \href {https://doi.org/10.1103/PhysRevA.37.935}
  {\path{doi:10.1103/PhysRevA.37.935}}.

\bibitem{Georgelin2007}
M.~Georgelin, S.~Bodea, A.~Pocheau, Coherence of dendritic sidebranching in
  directional solidification, EPL 77~(4) (2007) 46001.
\newblock \href {https://doi.org/10.1209/0295-5075/77/46001}
  {\path{doi:10.1209/0295-5075/77/46001}}.

\bibitem{Echebarria2010}
B.~Echebarria, A.~Karma, S.~Gurevich, Onset of sidebranching in directional
  solidification, Phys. Rev. E 81~(2) (2010) 021608.
\newblock \href {https://doi.org/10.1103/PhysRevE.81.021608}
  {\path{doi:10.1103/PhysRevE.81.021608}}.

\bibitem{Gonzalez-Cinca2001}
R.~{Gonz\'alez-Cinca}, L.~{Ram\'irez-Piscina}, J.~Casademunt,
  A.~{Hern\'andez-Machado}, Sidebranching induced by external noise in solutal
  dendritic growth, Phys. Rev. E 63~(5) (2001) 051602.
\newblock \href {https://doi.org/10.1103/PhysRevE.63.051602}
  {\path{doi:10.1103/PhysRevE.63.051602}}.

\bibitem{Boettinger2002}
W.~J. Boettinger, J.~A. Warren, C.~Beckermann, A.~Karma, {Phase}-{Field}
  {Simulation} {of} {Solidification}, Ann. Rev. Materials Research 32~(1)
  (2002) 163--194.
\newblock \href {https://doi.org/10.1146/annurev.matsci.32.101901.155803}
  {\path{doi:10.1146/annurev.matsci.32.101901.155803}}.

\bibitem{Chen2002}
L.-Q. Chen, {Phase}-{Field} {Models} {for} {Microstructure} {Evolution}, Ann.
  Rev. Materials Research 32~(1) (2002) 113--140.
\newblock \href {https://doi.org/10.1146/annurev.matsci.32.112001.132041}
  {\path{doi:10.1146/annurev.matsci.32.112001.132041}}.

\bibitem{Pavlik2000}
S.~G. Pavlik, R.~F. Sekerka, Fluctuations in the phase-field model of
  solidification, Physica A 277~(3) (2000) 415--431.
\newblock \href {https://doi.org/10.1016/S0378-4371(99)00382-9}
  {\path{doi:10.1016/S0378-4371(99)00382-9}}.

\bibitem{Tong2001}
X.~Tong, C.~Beckermann, A.~Karma, Q.~Li, Phase-field simulations of dendritic
  crystal growth in a forced flow, Phys. Rev. E 63~(6) (2001) 061601.
\newblock \href {https://doi.org/10.1103/PhysRevE.63.061601}
  {\path{doi:10.1103/PhysRevE.63.061601}}.

\bibitem{Yamanaka2011}
A.~Yamanaka, T.~Aoki, S.~Ogawa, T.~Takaki, {GPU}-accelerated phase-field
  simulation of dendritic solidification in a binary alloy, J. Cryst. Growth
  318~(1) (2011) 40--45.
\newblock \href {https://doi.org/10.1016/j.jcrysgro.2010.10.096}
  {\path{doi:10.1016/j.jcrysgro.2010.10.096}}.

\bibitem{Lin2011}
H.~K. Lin, C.~C. Chen, C.~W. Lan, Adaptive three-dimensional phase-field
  modeling of dendritic crystal growth with high anisotropy, J. Cryst. Growth
  318~(1) (2011) 51--54.
\newblock \href {https://doi.org/10.1016/j.jcrysgro.2010.11.013}
  {\path{doi:10.1016/j.jcrysgro.2010.11.013}}.

\bibitem{Hurlimann1992}
E.~Hurlimann, R.~Trittibach, U.~Bisang, J.~Bilgram, Integral {{Parameters}} of
  {{Xenon Dendrites}}, Phys. Rev. A 46~(10) (1992) 6579--6595.
\newblock \href {https://doi.org/10.1103/PhysRevA.46.6579}
  {\path{doi:10.1103/PhysRevA.46.6579}}.

\bibitem{Bilgram1993a}
J.~Bilgram, The {{Structure}} and {{Properties}} of {{Melt}} and
  {{Concentrated}}-{{Solutions}}, Prog. Cryst. Growth Charact. Mater. 26 (1993)
  99--119.
\newblock \href {https://doi.org/10.1016/0960-8974(93)90012-S}
  {\path{doi:10.1016/0960-8974(93)90012-S}}.

\bibitem{Singer2006}
H.~M. Singer, J.~H. Bilgram, Integral scaling behavior of different
  morphologies of {{3D}} xenon crystals, Physica D: Nonlinear Phenomena 219~(2)
  (2006) 101--110.
\newblock \href {https://doi.org/10.1016/j.physd.2006.05.018}
  {\path{doi:10.1016/j.physd.2006.05.018}}.

\bibitem{Dougherty2019a}
A.~J. Dougherty, Measurement of the capillary length for the dendritic growth
  of ammonium chloride, J. Cryst. Growth 506 (2019) 156--159.
\newblock \href {https://doi.org/10.1016/j.jcrysgro.2018.10.025}
  {\path{doi:10.1016/j.jcrysgro.2018.10.025}}.

\bibitem{AmmoniumChloride2018}
M.~{Stefan-Kharicha}, A.~Kharicha, J.~Mogeritsch, M.~Wu, A.~Ludwig, Review of
  {{Ammonium Chloride}}\textendash{{Water Solution Properties}}, J. Chem. Eng.
  Data 63~(9) (2018) 3170--3183.
\newblock \href {https://doi.org/10.1021/acs.jced.7b01062}
  {\path{doi:10.1021/acs.jced.7b01062}}.

\end{thebibliography}





\end{document}